\documentclass[print,superscriptaddress,twocolumn,10pt]{revtex4-1}
\usepackage{amsmath}
\usepackage{epsfig}
\usepackage{subfigure,mathrsfs}
\usepackage{array}
\usepackage{amssymb}
\usepackage[dvipsnames]{xcolor}

\newcommand{\beq}[0]{\begin{equation}}
\newcommand{\eeq}[0]{\end{equation}}
\def\be{\begin{equation}}
\def\ee{\end{equation}}
\def\bea{\begin{eqnarray}}
\def\eea{\end{eqnarray}}
\usepackage[breaklinks=true,colorlinks]{hyperref}

\newcommand{\kB}{k_\mathrm{B}}
\newcommand{\bop}{\mathsf{b}}

\begin{document}

\title{Two-level masers as heat-to-work converters}

\author{A. Ghosh}
\affiliation{Department of Chemical and Biological Physics, Weizmann Institute of Science, 
Rehovot 7610001, Israel}
\affiliation{Department of Physics, Shanghai University, Baoshan District, Shanghai 200444, P.R. China}
\author{D. Gelbwaser-Klimovsky}
\affiliation{Harvard University, USA}
\author{W. Niedenzu}
\affiliation{Department of Chemical and Biological Physics, Weizmann Institute of Science, 
Rehovot 7610001, Israel}
\author{A. Lvovsky}
\affiliation{Institute for Quantum Science and Technology, University of Calgary, Calgary, Alberta,
	Canada T2N 1N4}
\affiliation{Russian Quantum Center, 100 Novaya St., Skolkovo, Moscow 143025, Russia}
\affiliation{Institute of Fundamental and Frontier Sciences, University of Electronic
	Science and Technology, Chengdu, Sichuan 610054, China}
\author{I. Mazets}
\affiliation{Atominstitut, TU Wien, Stadionallee 2, 1020 Vienna, Austria }
\affiliation{Wolfgang Pauli Institute c/o Fakult\"at f\"ur Mathematik, Universit\"at Wien, Oskar-Morgenstern-Platz 1, 1090 Vienna, Austria}
\author{M. O. Scully}
\affiliation{Texas A $\&$ M University, USA}
\affiliation{Princeton University, Princeton, NJ 08544 }
\affiliation{Baylor University, Waco, TX ,76798}
\author{G. Kurizki}
\affiliation{Department of Chemical and Biological Physics, Weizmann Institute of Science, 
Rehovot 7610001, Israel}

\begin{abstract}
Heat engines, which cyclically transform heat into work, are ubiquitous in technology. Lasers  and masers, which generate a  coherent electromagnetic field, may be viewed  as heat engines that rely on population inversion or coherence in the active medium. Here we put forward an unconventional paradigm of a remarkably simple electromagnetic  heat-powered engine that bears basic differences to any known maser or laser: it does not rely on population inversion or coherence in its \textit{two-level} working medium. Nor does it require any coherent driving or pump aside from two (hot and cold) baths. Strikingly, the proposed maser, in which the heat exchange between these baths mediated by the working medium amplifies the signal field, can attain the highest possible efficiency even if the signal is incoherent.  
\end{abstract}
\maketitle

\section{Introduction}

\begin{figure}
\centering
\includegraphics[width=\columnwidth]{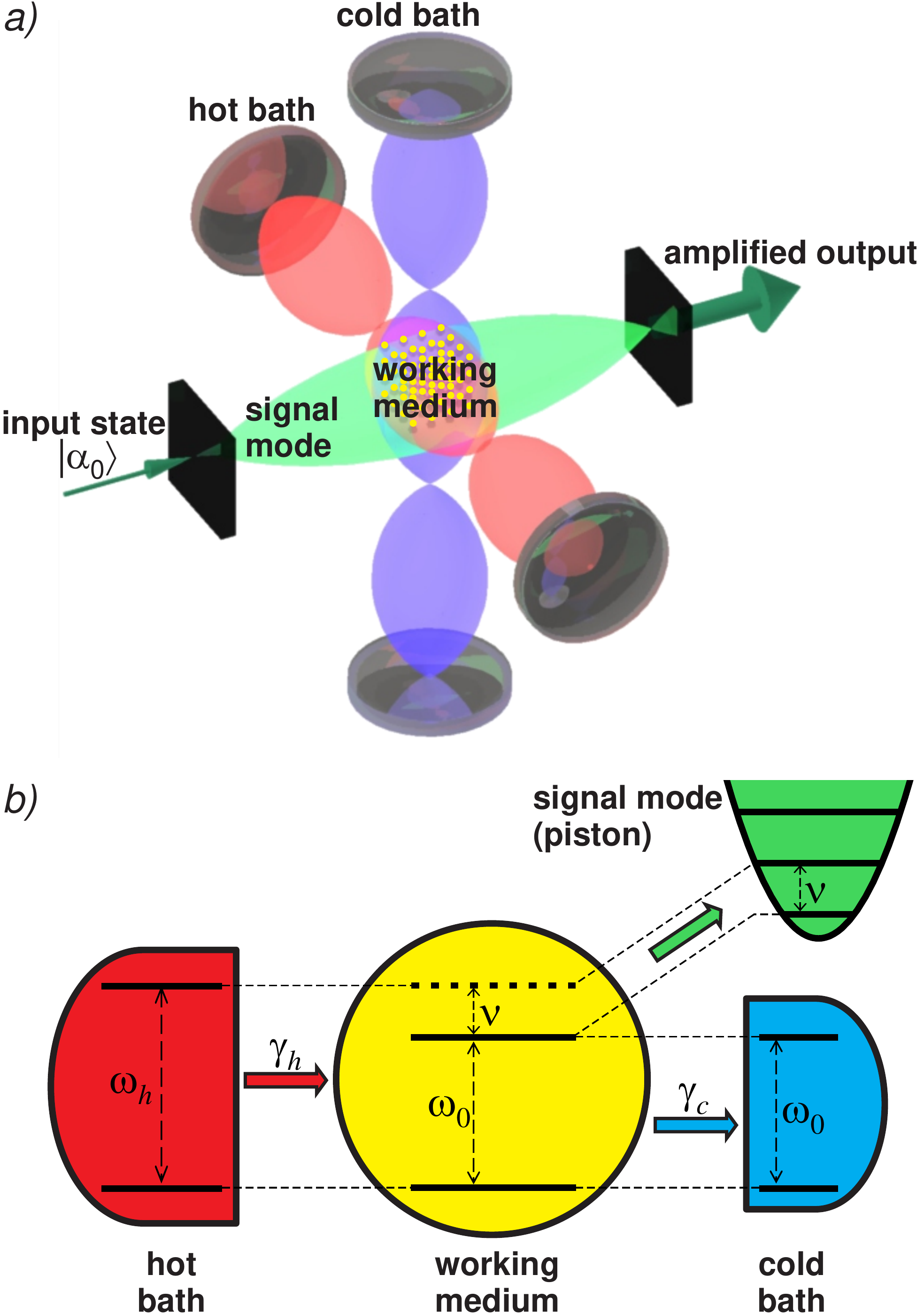}
\caption{a) The proposed  quantum heat engine acting as a maser. The two narrow-band cavities spectrally filter the hot and cold baths. The working medium consists of TLSs with energy-level distance $\omega_{0}$ and an off-resonant microwave or infrared signal (green) at frequency $\nu$. b) A thermodynamic outline of the proposed scheme.} 
\label{fig:1}
\end{figure} 
A maser or a laser is typically based on a medium with three or more levels wherein one transition is pumped to establish population inversion on another (signal) transition \cite{Bertolotti_BOOK}. Alternatively, pumping may serve to induce coherence between a set of levels, as in the case of lasers without inversion (LWI) \cite{Scully_BOOK}. Either scheme results in the pump conversion into a \textit{coherent}, amplified signal output.

\par

A much less emphasized aspect of lasers and masers is their analogy to heat engines. This analogy was pointed out by Scovil and Schulz-DuBois (SSD) who related \cite{Scovil_1959} the maximal conversion efficiency of a three-level maser to the Carnot efficiency bound \cite{Callen_BOOK} by assuming that one transition is pumped by a hot bath and another is coupled to a cold bath. The three-level SSD scheme has become a canonical microscopic model for heat engines \cite{Boukobza_2007,Kosloff_2014} and their quantum-mechanical characteristics \cite{Gelbwaser_2013_b,Scully_2011,Uzdin_2015}.

\par

In view of the extensive developments in the area of quantum heat engines \cite{Scully_2003,Rio_2011,Skrzypczyk_2014,Pekola_2015,Gelbwaser_2015,Lostaglio_2015,Brandao_2015,Rossnagel_2016,Vinjanampathy_2016,
niedenzu2017universal,Kosloff_2017,Klatzow_2017}, we find it timely to revisit the rapport between masers and heat engines. To this end, we put forward an unconventional operational paradigm of a remarkably-simple, heat-powered maser. Against the common view that at least two transitions are needed for a maser/laser, the proposed device only employs one (two-photon, i.e. Raman) transition and its working medium (WM) consists of \textit{uninverted two-level systems} (TLSs). We identify the  work \cite{Pusz_1978,Lenard_1978,Allahverdyan_2004}  
output of the proposed device and use it to infer the thermodynamic efficiency. We find that neither the input nor the output need be in a coherent state to achieve the highest (SSD) efficiency.

\par

The basic ingredients of the envisaged device are [Fig.~\ref{fig:1}a]: (i) hot and cold heat baths, realized by ambient thermal radiation that is filtered into spectrally distinct modes of (infrared or microwave) narrow-band cavities; (ii) a WM consisting of TLSs that are continuously \cite{Kosloff_2014,Gelbwaser_2015} (rather than intermittently, as in Otto or Carnot cycles \cite{Scully_2003,Rossnagel_2016,Kosloff_2017}) coupled to the hot and cold baths (cavity modes); (iii) a microwave or infrared-signal mode which acts as a ``piston'' that extracts the work.

\par

The two levels of the WM are coupled to the cold bath near a frequency $\omega_{0}$ and (via a two-photon transition) to the hot bath, near the frequency  $\omega_h$ as well as to the signal near  $\nu =\omega_h -\omega_{0}$. The heat-to-work conversion consists in photon absorption from the hot bath, re-emission into the cold bath and production of a signal photon at the difference frequency.

\par

Because the WM interacts with the cold bath resonantly, and hence more strongly, than with the hot bath, it can be assumed to be close to the temperature of the former. It is then the \textit{thermal imbalance} between the hot bath and the WM can play the role of a population-inverted two-level system for the signal, and lead to maser-like amplification. This becomes possible because the energy exchange between the hot bath and the WM is mediated by the signal mode.

\section{Operational Principle} 
We consider the WM whose $|e\rangle \leftrightarrow |g\rangle$ transition at frequency $\omega_0$ is near-resonant with a cold bath (c), and is in a two-photon (Raman) resonance with the signal mode at frequency $\nu$ and the hot-bath (h) modes near frequency $\omega_h$ [Fig.~\ref{fig:1}b]. 
The effective Hamiltonians of the interaction between the WM and the two baths in the interaction picture and the rotating-wave approximation (RWA) are $V_c(t)$ and $V_h(t)$ respectively. They are written in terms of $a^{(c,h)}_{k}$ and ${\bop}^{\dag}$, the mode annihilation and creation operators of the two baths and signal, respectively. While the Hamiltonian $V_c(t)$ for resonant system-bath coupling comprised of $|g\rangle\langle e|a^{(c)}_{k}$ terms is well known \cite{Scully_BOOK}, the Raman coupling Hamiltonian $V_h(t)$ consisting of $|g\rangle\langle e|{\bop}^\dag a^{(h)}_{k}$ is derived in Appendix~\ref{Suppl-A}.

\par

Because, as mentioned above, the WM interacts mainly with the near-resonant cold bath at temperature $T_c$ (i.e.~$V_h(t)$ has much weaker effect than $V_c(t)$), it attains a steady state whose upper-and lower-level populations $\rho_{ee}$ and $\rho_{gg}$ are related by 
\begin{gather}
\frac{\rho_{ee}}{\rho_{gg}} \simeq \exp\left(-\frac{\hbar\omega_0}{{\kB}T_c}\right)=\frac{\bar{n}_c}{\bar{n}_c+1},
  \label{population-ratio}
\end{gather} 
where 
\begin{equation}
\label{occupancies}\bar{n}_{c,h}=\left[\exp\left(\frac{\hbar\omega_{0,h}}{{\kB}T_{c,h}}\right)-1\right]^{-1}
\end{equation} 
are the photon occupancies of the two baths. The knowledge of the WM state now allows us to find the evolution of the signal field as a mediator of the interaction between the system and the hot bath that obeys the master equation (Appendix~\ref{Suppl-B})
\begin{gather}
\dot{\rho}_s=\gamma_h(\bar{n}_{h}+1)\rho_{ee}([{\bop}\rho_s,{\bop}^{\dag}]+[{\bop},\rho_s {\bop}^{\dag}])\nonumber\\
+\gamma_h\bar{n}_{h}\rho_{gg}([{\bop}^{\dag}\rho_s,{\bop}]+[{\bop}^{\dag},\rho_s {\bop}]),\label{rho-s-dot0}
\end{gather}
where $\gamma_h$ is the decay rate into the hot bath associated with the Raman coupling constant $g^{2}_{k}$.

\par

From this master equation, upon treating the signal mode \textit{semiclassically} (see Appendix~\ref{Suppl-B}), one can obtain a rate equation for the signal-mode intensity (mean energy) as 
\begin{gather}
\dot{I}_s=GI_s\nonumber\\
G=\gamma_h\left[\rho_{gg}\bar{n}_h-\rho_{ee}(\bar{n}_{h}+1)\right], 
\label{rateeq-scully}
\end{gather}
where $G$ is the gain of our maser. The factor in the square brackets of Eq.~(\ref{rateeq-scully}) expresses the difference between the photon stimulated-emission probability $\rho_{gg}\bar{n}_h$ and its absorption $\rho_{ee}(\bar{n}_{h}+1)$ induced by the hot bath.

\par

The gain (\ref{rateeq-scully}) can be rewritten with the help of Eq.~(\ref{population-ratio}) as 
\begin{gather}
	G=\gamma_h\frac{\bar{n}_h-\bar{n}_c}{2\bar{n}_c+1}. 
	\label{rateeq-scully-mod}
\end{gather}
This means that $G \geq 0$, above or at the amplification threshold, corresponds to $\bar{n}_{h}(\omega_h) \geq \bar{n}_{c}(\omega_0)$ and hence to
\begin{equation}\label{threshold}
\frac{\omega_h}{T_h} \leq \frac{\omega_c}{T_c}.
\end{equation}

\par

The energy $\hbar\omega_h$ of a hot-bath photon is shared between the signal-mode and cold-bath photons, with energies $\hbar\nu$ and $\hbar\omega_c$, respectively. This sets the limit on the efficiency of our engine, defined as the ratio of the extracted work in the signal mode to the heat input from the hot bath.

\par

Since $\omega_c=\omega_h-\nu$, the maser efficiency [conforming to~\eqref{rateeq-scully}--\eqref{threshold}] satisfies
\begin{gather}
\eta=\eta_{\mathrm{SSD}}=\frac{\nu}{\omega_h} \leq 1-\frac{T_c}{T_h},
\label{Carnot-efficiency}
\end{gather}
where the equality holds at threshold in the semiclassical regime for the signal. Namely, we recover the SSD relation \cite{Scovil_1959} for the engine efficiency at the threshold point whereby the signal-to the hot-pump frequency ratio corresponds to the Carnot limit. Below threshold, i.e., for $G<0$, the device acts as a \textit{refrigerator}: it consumes the signal power in order to move heat from the cold to the hot baths \cite{Boukobza_2007}.

\par

In contrast to the SSD model for a maser, here the WM is uninverted and no other transition (level) is involved. Thermal-occupancy imbalance $\bar{n}_{h}(\omega_h)>\bar{n}_{c}(\omega_0)$ at two different frequencies, $\omega_h$ and $\omega_c$, rather than population inversion $\rho_{ee} > \rho_{gg}$ in a conventional (SSD) maser, is the requirement for the output signal intensity to exceed its input counterpart.

\section{Quantized Photonic Work}
\begin{figure}
	\centering
        \includegraphics[width=\columnwidth]{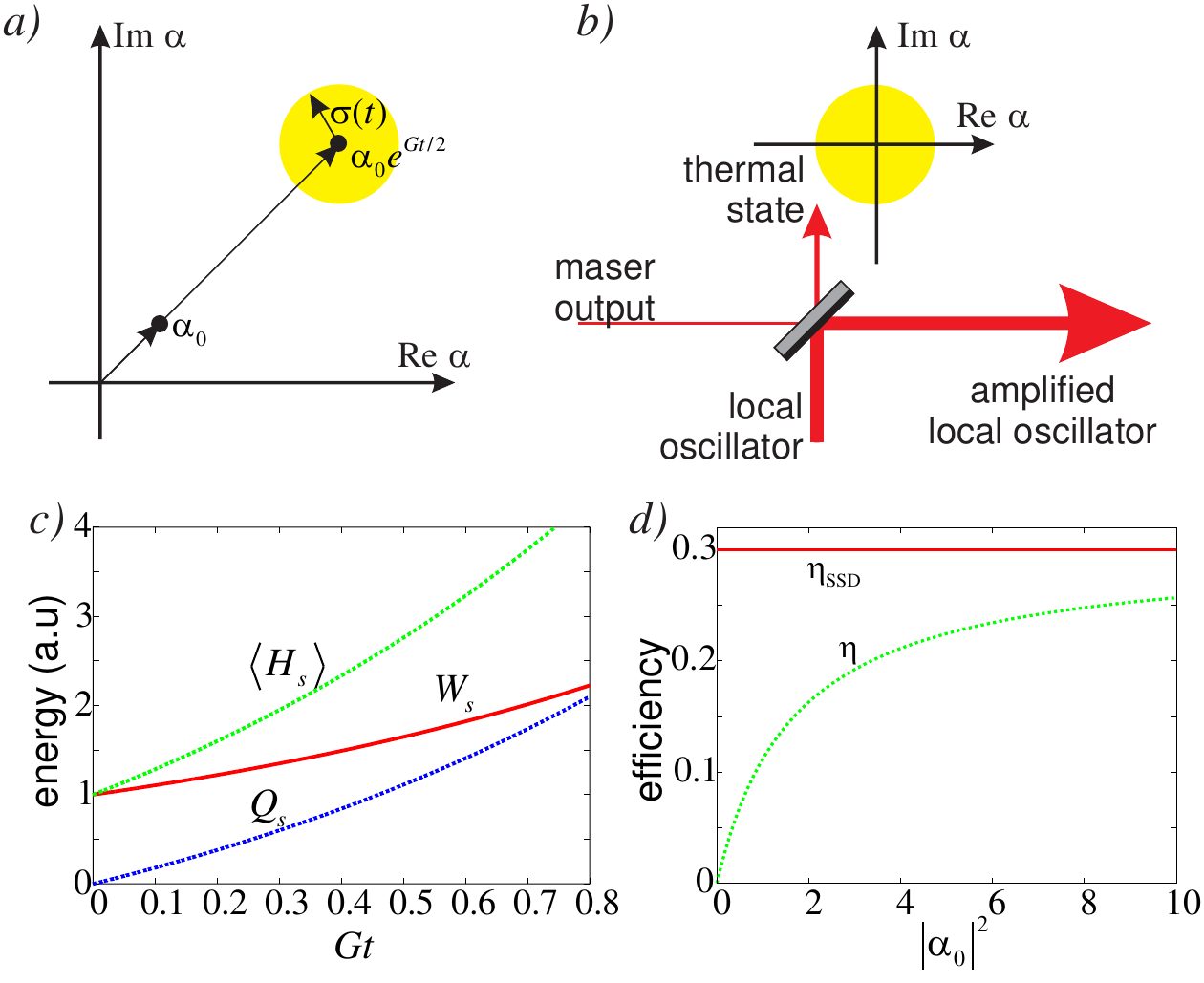}
\caption{Amplification of the signal state. a) The Glauber-Sudarshan $P$ function of an initial coherent state $|\alpha_0\rangle$ and the amplified state. The initial $P$ function is singular while the final state is a displaced thermal state, so it has a finite width $\sigma(t)$. b) Separation of the active (work-producing) and passive (heat-producing) components of the signal output by means of the displacement operator implemented via a beam splitter (BS) transformation. c) Small signal amplification: $Q_s$, the passive (thermal) energy, ergotropy or work $\mathcal{W}_s$ and total mean energy $\langle H_s \rangle$ for initial $|\alpha_{0}|^{2}=1$. d) Efficiency $\eta$ (work output divided by the heat input) as a function of the mean squared amplitude $|\alpha_{0}|^{2}$ tends to the semiclassical efficiency $\eta_\mathrm{SSD}=\nu/\omega_{h}$ limited by the Carnot bound for large initial-state amplitudes $|\alpha_0|$. The parameters for Figs. (c) and (d) are: $\nu/\omega_h=0.3$, $\gamma_h /\nu=0.033$ and $T_h=2.33 T_c$.
}
\label{fig:2}
\end{figure}
The above discussion has assumed, as is common for heat machines \cite{Callen_BOOK,Kosloff_2014,Gelbwaser_2015}, 
that the piston (here the signal mode) is semiclassical and its entropy is negligible (hence it is often referred to as work reservoir). However, in the case of a microscopic input signal, we must allow for the ``heating up'' of the signal mode \cite{Gelbwaser_2013_b}. That is, part of the energy acquired by the piston leads to its entropy increase and cannot be extracted as useful work.

\par

To find the maximal extractable work (alas ergotropy) \cite{Pusz_1978,Lenard_1978,Allahverdyan_2004} of the signal mode, a fully quantum treatment of this mode is required. To this end we consider the evolution of the Glauber-Sudarshan $P$-function of the signal mode~\cite{Scully_BOOK}, governed by a Fokker-Planck equation (Appendix~\ref{Suppl-B}). For an initial coherent state $|\alpha_0\rangle$ we find that the distribution (in the interaction picture) evolves as
		\begin{equation}\label{Pevolve}
		P(\alpha,t)=\frac1{\pi\sigma^2(t)}\exp\left[-\frac{|\alpha-\alpha_0 e^{G t/2}|^2}{\sigma^2(t)}\right].
		\end{equation}
		Thus the distribution is shifted outward by $\alpha_0 e^{G t/2}$ and its width grows as
		\begin{equation}\label{sigma}
		\sigma^2(t)=\frac{\bar{n}_h(\bar{n}_{c}+1)}{\bar{n}_{h}-\bar{n}_{c}}(e^{G t}-1).
		\end{equation}
That is, a coherent state at the input becomes a \textit{displaced} \emph{thermal} state [Fig.~\ref{fig:2}(a)]. The mean energy of this state is \cite{Leonhardt_BOOK}
\begin{align}
\langle H_s(t)\rangle&=\hbar\nu\left[|\alpha_{0}|^{2}e^{G t}+\sigma^2(t)\right],
\label{eq:Hstavg}
\end{align}
where the first term in the brackets corresponds to the coherent amplitude and the second to the thermal contribution.

\par

For a given signal state $\rho_s$, the ergotropy \cite{Pusz_1978,Lenard_1978,Allahverdyan_2004,niedenzu2017universal}, which is the maximal extractable work from any state, is expressed by
	\begin{equation}
	\mathcal{W} (\rho_s)= \langle H(\rho_s) \rangle - \langle H(\rho^\mathrm{pas}_{s})\rangle.
	\label{eq:maxw}
	\end{equation}
Here $\rho^\mathrm{pas}_{s}$ is a  \textit{passive state:} it is defined as the state with the least energy that is accessible from $\rho_s$ via a unitary transformation. Namely, Eq. \eqref{eq:maxw} expresses the maximal energy change that can be effected \textit{without entropy change} (isentropically), which is the definition of work extractable from the signal while keeping the Hamiltonian constant \cite{niedenzu2017universal}.

\par

In the case of the displaced thermal state, the second term in the expression (\ref{eq:Hstavg}) for the energy corresponds to a passive state. Indeed, the phase-space displacement operator $\mathcal{D}(-\alpha)=e^{-\alpha a^\dag+\alpha^*a}$, for $\alpha=\alpha_0e^{G t/2}$ applied to the amplified signal mode at $t$ will transform it to a thermal state centered at the phase space origin. This operation is unitary, and hence isentropic. The resulting thermal state of energy $\hbar\nu\sigma^2(t)$, on the other hand, is entirely passive \cite{Skrzypczyk_2015} 
and permits no further work extraction. Hence the ergotropy of the output state is $\mathcal{W} (t)=\hbar\nu|\alpha_{0}|^{2}e^{G t}$.

\par

In practice, the displacement operator can be realized by overlapping the target state on a highly reflective beam splitter \cite{Paris_1996,Lvovsky_2002} with a strong coherent local oscillator field  [Fig.~\ref{fig:2}(b) and Appendix~\ref{Suppl-C}]. A preliminary phase estimation on the input would be beneficial in order to adjust the local oscillator phase for the required displacement. Our work-extraction protocol depicted in Fig. \ref{fig:2}b assumes that the mean phase variance of the signal is (roughly below 1) not excessive, so that it can properly interfere with thermal oscillator.

\par

The efficiency of the engine with a coherent state $|\alpha_0\rangle$ at the input is found to be (Appendix~\ref{Suppl-D})
\begin{equation}\label{eff}
\eta=\eta_\mathrm{SSD}\frac{\mathcal{W} (t)-\mathcal{W} (0)}{\langle H_s(t)\rangle-\langle H_s(0)\rangle}=\frac{\nu}{\omega_{h}}\frac{|\alpha_{0}|^{2}}{|\alpha_{0}|^{2}+\frac{\bar{n}_h(\bar{n}_{c}+1)}{\bar{n}_{h}-\bar{n}_{c}}}.
\end{equation}
We see that the efficiency obtained by the quantum treatment is always less than its classical counterpart, but approaches it for high input-state amplitudes [Fig. \ref{fig:2}(c,d)].

\par

If the engine is \textit{repeatedly} operated by injecting independent coherent states whose amplitudes are distributed according to $p(|\alpha_0|)$, the upper bound on the mean efficiency can be calculated to be
\begin{gather}
\langle\eta\rangle = \frac{\nu}{\omega_{h}}\int \frac{|\alpha_{0}|^{2}}{|\alpha_{0}|^{2}+\frac{\bar{n}_h(\bar{n}_{c}+1)}{\bar{n}_{h}-\bar{n}_{c}}} |\alpha_0|p(|\alpha_0|)\mathrm{d}|\alpha_0| \leqslant \eta_{\mathrm{SSD}}.
\label{eq:efficiency-final}
\end{gather}
We see that efficiencies up to the SSD limit are obtainable provided the mean initial amplitude $\langle|\alpha_{0}|\rangle$ dominates over the thermal term $\bar{n}_h(\bar{n}_{c}+1)/(\bar{n}_{h}-\bar{n}_{c})$.

\section{Proposed experimental setups} 
The envisioned experiments may be conducted with thermal atoms or molecules in microwave cavities \cite{Scully_BOOK}. One cavity mode can be in contact with solar radiation and the other with the ground (earth surface) (Fig.~\ref{fig:1}). The WM may consist of, e.g., pentacene molecules in a host lattice at room temperature, with a transition (between spin triplet-sublevels) at 1.45 GHz. Whereas optical pumping is normally required for population inversion \cite{Oxborrow_2012,Singer_BOOK}, it is not needed in the proposed scheme: ambient heat may be converted to work at the signal frequency $\nu \gtrsim 1$ MHz: for Raman coupling $g_k \gtrsim 500$ kHz, we obtain a gain rate $G \gtrsim 100$ kHz.

\par

In our quest for an optimized WM, we may consider a transition with giant transition dipole between two Rydberg levels with $\omega_{0}$ in the GHz range. The downside of this implementation is that it requires pumping of the lower level in the Rydberg transition. The upside is that the power output from a Rydberg atom $^{87}$Rb in the level $n=68$ may be higher by 3 orders of magnitude (\textit{per atom}) than in Ref. \cite{Rossnagel_2016} (where the power per particle (ion) is $3.4\times 10^{-24}$ W).

\section{Conclusions} 
The analysis of the proposed schemes indicates the experimental feasibility of a hitherto unexplored principle: a two-level maser powered by heat, and its ability to act as a photonic heat engine. The scheme is \textit{universal}: we can always choose a transition at frequency $\omega_0$ to be resonant with the cold-bath cavity mode and tune the signal to the frequency difference $\nu$ between the hot-and cold cavity modes, so that the signal is amplified at the expense of the difference between thermal-quanta occupancies of the two cavity modes.

\par

We have studied the effect of coherence in the input and output states on the extractable work. The required property of the signal-mode output is non-passivity \cite{Gelbwaser_2013_b}. As long as the initial mean phase variance is not excessive during repeated operations of the engine, work extraction is straightforward by mixing the output signal with a local oscillator. Yet their phase spread may be much larger than that of the relevant coherent state.

\par

From the technological point of view the ability to extract useful photonic work (power) from ambient heat at the temperature difference between, say, the atmosphere and the ground surface is enticing: the fuel of this engine is ``free for the taking''; the engine is simple and robust and most importantly, being autonomous (self-contained), directly pumped by heat. Hence, it comes as close to the ``perfect engine'' dream as is allowed by the laws of thermodynamics (that yield the Carnot bound). 

\par

\textbf{Acknowledgement:} We acknowledge the support of the US-Israel BSF (G.K. and M.O.S.), the ISF (G.K), NSERC (A.L.), the Austrian Science Foundation (FWF) (project P 25329-N27) (I.M.). A.L. is a CIFAR Fellow.


\begin{thebibliography}{33}%
\makeatletter
\providecommand \@ifxundefined [1]{%
 \@ifx{#1\undefined}
}%
\providecommand \@ifnum [1]{%
 \ifnum #1\expandafter \@firstoftwo
 \else \expandafter \@secondoftwo
 \fi
}%
\providecommand \@ifx [1]{%
 \ifx #1\expandafter \@firstoftwo
 \else \expandafter \@secondoftwo
 \fi
}%
\providecommand \natexlab [1]{#1}%
\providecommand \enquote  [1]{``#1''}%
\providecommand \bibnamefont  [1]{#1}%
\providecommand \bibfnamefont [1]{#1}%
\providecommand \citenamefont [1]{#1}%
\providecommand \href@noop [0]{\@secondoftwo}%
\providecommand \href [0]{\begingroup \@sanitize@url \@href}%
\providecommand \@href[1]{\@@startlink{#1}\@@href}%
\providecommand \@@href[1]{\endgroup#1\@@endlink}%
\providecommand \@sanitize@url [0]{\catcode `\\12\catcode `\$12\catcode
  `\&12\catcode `\#12\catcode `\^12\catcode `\_12\catcode `\%12\relax}%
\providecommand \@@startlink[1]{}%
\providecommand \@@endlink[0]{}%
\providecommand \url  [0]{\begingroup\@sanitize@url \@url }%
\providecommand \@url [1]{\endgroup\@href {#1}{\urlprefix }}%
\providecommand \urlprefix  [0]{URL }%
\providecommand \Eprint [0]{\href }%
\providecommand \doibase [0]{http://dx.doi.org/}%
\providecommand \selectlanguage [0]{\@gobble}%
\providecommand \bibinfo  [0]{\@secondoftwo}%
\providecommand \bibfield  [0]{\@secondoftwo}%
\providecommand \translation [1]{[#1]}%
\providecommand \BibitemOpen [0]{}%
\providecommand \bibitemStop [0]{}%
\providecommand \bibitemNoStop [0]{.\EOS\space}%
\providecommand \EOS [0]{\spacefactor3000\relax}%
\providecommand \BibitemShut  [1]{\csname bibitem#1\endcsname}%
\let\auto@bib@innerbib\@empty
\bibitem [{\citenamefont {Bertolotti}(1983)}]{Bertolotti_BOOK}%
  \BibitemOpen
  \bibfield  {author} {\bibinfo {author} {\bibfnamefont {M.}~\bibnamefont
  {Bertolotti}},\ }\href@noop {} {\emph {\bibinfo {title} {Masers and Lasers:
  An Historical Approach}}}\ (\bibinfo  {publisher} {CRC Press},\ \bibinfo
  {address} {New York, US},\ \bibinfo {year} {1983})\BibitemShut {NoStop}%
\bibitem [{\citenamefont {Scully}\ and\ \citenamefont
  {Zubairy}(1997)}]{Scully_BOOK}%
  \BibitemOpen
  \bibfield  {author} {\bibinfo {author} {\bibfnamefont {M.~O.}\ \bibnamefont
  {Scully}}\ and\ \bibinfo {author} {\bibfnamefont {M.~S.}\ \bibnamefont
  {Zubairy}},\ }\href@noop {} {\emph {\bibinfo {title} {Quantum Optics}}}\
  (\bibinfo  {publisher} {Cambridge University Press},\ \bibinfo {address}
  {Cambridge, UK},\ \bibinfo {year} {1997})\BibitemShut {NoStop}%
\bibitem [{\citenamefont {Scovil}\ and\ \citenamefont
  {Schulz-DuBois}(1959)}]{Scovil_1959}%
  \BibitemOpen
  \bibfield  {author} {\bibinfo {author} {\bibfnamefont {H.~E.~D.}\
  \bibnamefont {Scovil}}\ and\ \bibinfo {author} {\bibfnamefont {E.~O.}\
  \bibnamefont {Schulz-DuBois}},\ }\href {\doibase 10.1103/PhysRevLett.2.262}
  {\bibfield  {journal} {\bibinfo  {journal} {Phys. Rev. Lett.}\ }\textbf
  {\bibinfo {volume} {2}},\ \bibinfo {pages} {262} (\bibinfo {year}
  {1959})}\BibitemShut {NoStop}%
\bibitem [{\citenamefont {Callen}(1985)}]{Callen_BOOK}%
  \BibitemOpen
  \bibfield  {author} {\bibinfo {author} {\bibfnamefont {H.~B.}\ \bibnamefont
  {Callen}},\ }\href@noop {} {\emph {\bibinfo {title} {Thermodynamics and an
  Introduction to Thermostatistics}}}\ (\bibinfo  {publisher} {John Wiley \&
  Son},\ \bibinfo {address} {Singapore},\ \bibinfo {year} {1985})\BibitemShut
  {NoStop}%
\bibitem [{\citenamefont {Boukobza}\ and\ \citenamefont
  {Tannor}(2007)}]{Boukobza_2007}%
  \BibitemOpen
  \bibfield  {author} {\bibinfo {author} {\bibfnamefont {E.}~\bibnamefont
  {Boukobza}}\ and\ \bibinfo {author} {\bibfnamefont {D.~J.}\ \bibnamefont
  {Tannor}},\ }\href {\doibase 10.1103/PhysRevLett.98.240601} {\bibfield
  {journal} {\bibinfo  {journal} {Phys. Rev. Lett.}\ }\textbf {\bibinfo
  {volume} {98}},\ \bibinfo {pages} {240601} (\bibinfo {year}
  {2007})}\BibitemShut {NoStop}%
\bibitem [{\citenamefont {Kosloff}\ and\ \citenamefont
  {Levy}(2014)}]{Kosloff_2014}%
  \BibitemOpen
  \bibfield  {author} {\bibinfo {author} {\bibfnamefont {R.}~\bibnamefont
  {Kosloff}}\ and\ \bibinfo {author} {\bibfnamefont {A.}~\bibnamefont {Levy}},\
  }\href {\doibase 10.1146/annurev-physchem-040513-103724} {\bibfield
  {journal} {\bibinfo  {journal} {Annu. Rev. Phys. Chem.}\ }\textbf {\bibinfo
  {volume} {65}},\ \bibinfo {pages} {365} (\bibinfo {year} {2014})}\BibitemShut
  {NoStop}%
\bibitem [{\citenamefont {Gelbwaser-Klimovsky}, \citenamefont {Alicki},\ and\
  \citenamefont {Kurizki}(2013)}]{Gelbwaser_2013_b}%
  \BibitemOpen
  \bibfield  {author} {\bibinfo {author} {\bibfnamefont {D.}~\bibnamefont
  {Gelbwaser-Klimovsky}}, \bibinfo {author} {\bibfnamefont {R.}~\bibnamefont
  {Alicki}}, \ and\ \bibinfo {author} {\bibfnamefont {G.}~\bibnamefont
  {Kurizki}},\ }\href {http://stacks.iop.org/0295-5075/103/i=6/a=60005}
  {\bibfield  {journal} {\bibinfo  {journal} {EPL (Europhysics Letters)}\
  }\textbf {\bibinfo {volume} {103}},\ \bibinfo {pages} {60005} (\bibinfo
  {year} {2013})}\BibitemShut {NoStop}%
\bibitem [{\citenamefont {Scully}\ \emph {et~al.}(2011)\citenamefont {Scully},
  \citenamefont {Chapin}, \citenamefont {Dorfman}, \citenamefont {Kim},\ and\
  \citenamefont {Svidzinsky}}]{Scully_2011}%
  \BibitemOpen
  \bibfield  {author} {\bibinfo {author} {\bibfnamefont {M.~O.}\ \bibnamefont
  {Scully}}, \bibinfo {author} {\bibfnamefont {K.~R.}\ \bibnamefont {Chapin}},
  \bibinfo {author} {\bibfnamefont {K.~E.}\ \bibnamefont {Dorfman}}, \bibinfo
  {author} {\bibfnamefont {M.~B.}\ \bibnamefont {Kim}}, \ and\ \bibinfo
  {author} {\bibfnamefont {A.}~\bibnamefont {Svidzinsky}},\ }\href {\doibase
  10.1073/pnas.1110234108} {\bibfield  {journal} {\bibinfo  {journal} {Proc.
  Natl. Acad. Sci.}\ }\textbf {\bibinfo {volume} {108}},\ \bibinfo {pages}
  {15097} (\bibinfo {year} {2011})}\BibitemShut {NoStop}%
\bibitem [{\citenamefont {Uzdin}, \citenamefont {Levy},\ and\ \citenamefont
  {Kosloff}(2015)}]{Uzdin_2015}%
  \BibitemOpen
  \bibfield  {author} {\bibinfo {author} {\bibfnamefont {R.}~\bibnamefont
  {Uzdin}}, \bibinfo {author} {\bibfnamefont {A.}~\bibnamefont {Levy}}, \ and\
  \bibinfo {author} {\bibfnamefont {R.}~\bibnamefont {Kosloff}},\ }\href
  {\doibase 10.1103/PhysRevX.5.031044} {\bibfield  {journal} {\bibinfo
  {journal} {Phys. Rev. X}\ }\textbf {\bibinfo {volume} {5}},\ \bibinfo {pages}
  {031044} (\bibinfo {year} {2015})}\BibitemShut {NoStop}%
\bibitem [{\citenamefont {Scully}\ \emph {et~al.}(2003)\citenamefont {Scully},
  \citenamefont {Zubairy}, \citenamefont {Agarwal},\ and\ \citenamefont
  {Walther}}]{Scully_2003}%
  \BibitemOpen
  \bibfield  {author} {\bibinfo {author} {\bibfnamefont {M.~O.}\ \bibnamefont
  {Scully}}, \bibinfo {author} {\bibfnamefont {M.~S.}\ \bibnamefont {Zubairy}},
  \bibinfo {author} {\bibfnamefont {G.~S.}\ \bibnamefont {Agarwal}}, \ and\
  \bibinfo {author} {\bibfnamefont {H.}~\bibnamefont {Walther}},\ }\href
  {\doibase 10.1126/science.1078955} {\bibfield  {journal} {\bibinfo  {journal}
  {Science}\ }\textbf {\bibinfo {volume} {299}},\ \bibinfo {pages} {862}
  (\bibinfo {year} {2003})}\BibitemShut {NoStop}%
\bibitem [{\citenamefont {Rio}\ \emph {et~al.}(2011)\citenamefont {Rio},
  \citenamefont {{\AA}berg}, \citenamefont {Renner}, \citenamefont {Dahlsten},\
  and\ \citenamefont {Vedral}}]{Rio_2011}%
  \BibitemOpen
  \bibfield  {author} {\bibinfo {author} {\bibfnamefont {L.~d.}\ \bibnamefont
  {Rio}}, \bibinfo {author} {\bibfnamefont {J.}~\bibnamefont {{\AA}berg}},
  \bibinfo {author} {\bibfnamefont {R.}~\bibnamefont {Renner}}, \bibinfo
  {author} {\bibfnamefont {O.}~\bibnamefont {Dahlsten}}, \ and\ \bibinfo
  {author} {\bibfnamefont {V.}~\bibnamefont {Vedral}},\ }\href
  {http://dx.doi.org/10.1038/nature10123} {\bibfield  {journal} {\bibinfo
  {journal} {Nature}\ }\textbf {\bibinfo {volume} {474}},\ \bibinfo {pages} {61
  EP } (\bibinfo {year} {2011})}\BibitemShut {NoStop}%
\bibitem [{\citenamefont {Skrzypczyk}, \citenamefont {Short},\ and\
  \citenamefont {Popescu}(2014)}]{Skrzypczyk_2014}%
  \BibitemOpen
  \bibfield  {author} {\bibinfo {author} {\bibfnamefont {P.}~\bibnamefont
  {Skrzypczyk}}, \bibinfo {author} {\bibfnamefont {A.~J.}\ \bibnamefont
  {Short}}, \ and\ \bibinfo {author} {\bibfnamefont {S.}~\bibnamefont
  {Popescu}},\ }\href {http://dx.doi.org/10.1038/ncomms5185} {\bibfield
  {journal} {\bibinfo  {journal} {Nat. Commun.}\ }\textbf {\bibinfo {volume}
  {5}} (\bibinfo {year} {2014})}\BibitemShut {NoStop}%
\bibitem [{\citenamefont {Pekola}(2015)}]{Pekola_2015}%
  \BibitemOpen
  \bibfield  {author} {\bibinfo {author} {\bibfnamefont {J.~P.}\ \bibnamefont
  {Pekola}},\ }\href {http://dx.doi.org/10.1038/nphys3169} {\bibfield
  {journal} {\bibinfo  {journal} {Nat Phys}\ }\textbf {\bibinfo {volume}
  {11}},\ \bibinfo {pages} {118} (\bibinfo {year} {2015})},\ \bibinfo {note}
  {progress Article}\BibitemShut {NoStop}%
\bibitem [{\citenamefont {Gelbwaser-Klimovsky}, \citenamefont {Niedenzu},\ and\
  \citenamefont {Kurizki}(2015)}]{Gelbwaser_2015}%
  \BibitemOpen
  \bibfield  {author} {\bibinfo {author} {\bibfnamefont {D.}~\bibnamefont
  {Gelbwaser-Klimovsky}}, \bibinfo {author} {\bibfnamefont {W.}~\bibnamefont
  {Niedenzu}}, \ and\ \bibinfo {author} {\bibfnamefont {G.}~\bibnamefont
  {Kurizki}},\ }\href {\doibase 10.1016/bs.aamop.2015.07.002} {\bibfield
  {journal} {\bibinfo  {journal} {Adv. At. Mol. Opt. Phys.}\ }\textbf {\bibinfo
  {volume} {64}},\ \bibinfo {pages} {329 } (\bibinfo {year}
  {2015})}\BibitemShut {NoStop}%
\bibitem [{\citenamefont {Lostaglio}, \citenamefont {Jennings},\ and\
  \citenamefont {Rudolph}(2015)}]{Lostaglio_2015}%
  \BibitemOpen
  \bibfield  {author} {\bibinfo {author} {\bibfnamefont {M.}~\bibnamefont
  {Lostaglio}}, \bibinfo {author} {\bibfnamefont {D.}~\bibnamefont {Jennings}},
  \ and\ \bibinfo {author} {\bibfnamefont {T.}~\bibnamefont {Rudolph}},\ }\href
  {\doibase 10.1038/ncomms7383} {\bibfield  {journal} {\bibinfo  {journal}
  {Nat. Commun.}\ }\textbf {\bibinfo {volume} {6}},\ \bibinfo {pages} {6383}
  (\bibinfo {year} {2015})}\BibitemShut {NoStop}%
\bibitem [{\citenamefont {Brandão}\ \emph {et~al.}(2015)\citenamefont
  {Brandão}, \citenamefont {Horodecki}, \citenamefont {Ng}, \citenamefont
  {Oppenheim},\ and\ \citenamefont {Wehner}}]{Brandao_2015}%
  \BibitemOpen
  \bibfield  {author} {\bibinfo {author} {\bibfnamefont {F.}~\bibnamefont
  {Brandão}}, \bibinfo {author} {\bibfnamefont {M.}~\bibnamefont {Horodecki}},
  \bibinfo {author} {\bibfnamefont {N.}~\bibnamefont {Ng}}, \bibinfo {author}
  {\bibfnamefont {J.}~\bibnamefont {Oppenheim}}, \ and\ \bibinfo {author}
  {\bibfnamefont {S.}~\bibnamefont {Wehner}},\ }\href {\doibase
  10.1073/pnas.1411728112} {\bibfield  {journal} {\bibinfo  {journal} {Proc.
  Natl. Acad. Sci.}\ }\textbf {\bibinfo {volume} {112}},\ \bibinfo {pages}
  {3275} (\bibinfo {year} {2015})}\BibitemShut {NoStop}%
\bibitem [{\citenamefont {Ro{\ss}nagel}\ \emph {et~al.}(2016)\citenamefont
  {Ro{\ss}nagel}, \citenamefont {Dawkins}, \citenamefont {Tolazzi},
  \citenamefont {Abah}, \citenamefont {Lutz}, \citenamefont {Schmidt-Kaler},\
  and\ \citenamefont {Singer}}]{Rossnagel_2016}%
  \BibitemOpen
  \bibfield  {author} {\bibinfo {author} {\bibfnamefont {J.}~\bibnamefont
  {Ro{\ss}nagel}}, \bibinfo {author} {\bibfnamefont {S.~T.}\ \bibnamefont
  {Dawkins}}, \bibinfo {author} {\bibfnamefont {K.~N.}\ \bibnamefont
  {Tolazzi}}, \bibinfo {author} {\bibfnamefont {O.}~\bibnamefont {Abah}},
  \bibinfo {author} {\bibfnamefont {E.}~\bibnamefont {Lutz}}, \bibinfo {author}
  {\bibfnamefont {F.}~\bibnamefont {Schmidt-Kaler}}, \ and\ \bibinfo {author}
  {\bibfnamefont {K.}~\bibnamefont {Singer}},\ }\href {\doibase
  10.1126/science.aad6320} {\bibfield  {journal} {\bibinfo  {journal}
  {Science}\ }\textbf {\bibinfo {volume} {352}},\ \bibinfo {pages} {325}
  (\bibinfo {year} {2016})}\BibitemShut {NoStop}%
\bibitem [{\citenamefont {Vinjanampathy}\ and\ \citenamefont
  {Anders}(2016)}]{Vinjanampathy_2016}%
  \BibitemOpen
  \bibfield  {author} {\bibinfo {author} {\bibfnamefont {S.}~\bibnamefont
  {Vinjanampathy}}\ and\ \bibinfo {author} {\bibfnamefont {J.}~\bibnamefont
  {Anders}},\ }\href {\doibase 10.1080/00107514.2016.1201896} {\bibfield
  {journal} {\bibinfo  {journal} {Contemp. Phys.}\ }\textbf {\bibinfo {volume}
  {57}},\ \bibinfo {pages} {545} (\bibinfo {year} {2016})}\BibitemShut
  {NoStop}%
\bibitem [{\citenamefont {Niedenzu}\ \emph {et~al.}(2017)\citenamefont
  {Niedenzu}, \citenamefont {Mukherjee}, \citenamefont {Ghosh}, \citenamefont
  {Kofman},\ and\ \citenamefont {Kurizki}}]{niedenzu2017universal}%
  \BibitemOpen
  \bibfield  {author} {\bibinfo {author} {\bibfnamefont {W.}~\bibnamefont
  {Niedenzu}}, \bibinfo {author} {\bibfnamefont {V.}~\bibnamefont {Mukherjee}},
  \bibinfo {author} {\bibfnamefont {A.}~\bibnamefont {Ghosh}}, \bibinfo
  {author} {\bibfnamefont {A.~G.}\ \bibnamefont {Kofman}}, \ and\ \bibinfo
  {author} {\bibfnamefont {G.}~\bibnamefont {Kurizki}},\ }\href
  {https://arxiv.org/abs/1703.02911} {\bibfield  {journal} {\bibinfo  {journal}
  {arXiv preprint arXiv:1703.02911}\ } (\bibinfo {year} {2017})}\BibitemShut
  {NoStop}%
\bibitem [{\citenamefont {Kosloff}\ and\ \citenamefont
  {Rezek}(2017)}]{Kosloff_2017}%
  \BibitemOpen
  \bibfield  {author} {\bibinfo {author} {\bibfnamefont {R.}~\bibnamefont
  {Kosloff}}\ and\ \bibinfo {author} {\bibfnamefont {Y.}~\bibnamefont
  {Rezek}},\ }\href {\doibase 10.3390/e19040136} {\bibfield  {journal}
  {\bibinfo  {journal} {Entropy}\ }\textbf {\bibinfo {volume} {19}},\ \bibinfo
  {pages} {136} (\bibinfo {year} {2017})}\BibitemShut {NoStop}%
\bibitem [{\citenamefont {Klatzow}\ \emph {et~al.}(2017)\citenamefont
  {Klatzow}, \citenamefont {Weinzetl}, \citenamefont {Ledingham}, \citenamefont
  {Becker}, \citenamefont {Saunders}, \citenamefont {Nunn}, \citenamefont
  {Walmsley}, \citenamefont {Uzdin},\ and\ \citenamefont
  {Poem}}]{Klatzow_2017}%
  \BibitemOpen
  \bibfield  {author} {\bibinfo {author} {\bibfnamefont {J.}~\bibnamefont
  {Klatzow}}, \bibinfo {author} {\bibfnamefont {C.}~\bibnamefont {Weinzetl}},
  \bibinfo {author} {\bibfnamefont {P.~M.}\ \bibnamefont {Ledingham}}, \bibinfo
  {author} {\bibfnamefont {J.~N.}\ \bibnamefont {Becker}}, \bibinfo {author}
  {\bibfnamefont {D.~J.}\ \bibnamefont {Saunders}}, \bibinfo {author}
  {\bibfnamefont {J.}~\bibnamefont {Nunn}}, \bibinfo {author} {\bibfnamefont
  {I.~A.}\ \bibnamefont {Walmsley}}, \bibinfo {author} {\bibfnamefont
  {R.}~\bibnamefont {Uzdin}}, \ and\ \bibinfo {author} {\bibfnamefont
  {E.}~\bibnamefont {Poem}},\ }\href {https://arxiv.org/abs/1710.08716}
  {\bibfield  {journal} {\bibinfo  {journal} {arXiv preprint arXiv:1710.08716}\
  } (\bibinfo {year} {2017})}\BibitemShut {NoStop}%
\bibitem [{\citenamefont {Pusz}\ and\ \citenamefont
  {Woronowicz}(1978)}]{Pusz_1978}%
  \BibitemOpen
  \bibfield  {author} {\bibinfo {author} {\bibfnamefont {W.}~\bibnamefont
  {Pusz}}\ and\ \bibinfo {author} {\bibfnamefont {S.~L.}\ \bibnamefont
  {Woronowicz}},\ }\href {\doibase 10.1007/BF01614224} {\bibfield  {journal}
  {\bibinfo  {journal} {Comm. Math. Phys.}\ }\textbf {\bibinfo {volume} {58}},\
  \bibinfo {pages} {273} (\bibinfo {year} {1978})}\BibitemShut {NoStop}%
\bibitem [{\citenamefont {Lenard}(1978)}]{Lenard_1978}%
  \BibitemOpen
  \bibfield  {author} {\bibinfo {author} {\bibfnamefont {A.}~\bibnamefont
  {Lenard}},\ }\href {\doibase 10.1007/BF01011769} {\bibfield  {journal}
  {\bibinfo  {journal} {J. Stat. Phys.}\ }\textbf {\bibinfo {volume} {19}},\
  \bibinfo {pages} {575} (\bibinfo {year} {1978})}\BibitemShut {NoStop}%
\bibitem [{\citenamefont {Allahverdyan}, \citenamefont {Balian},\ and\
  \citenamefont {Nieuwenhuizen}(2004)}]{Allahverdyan_2004}%
  \BibitemOpen
  \bibfield  {author} {\bibinfo {author} {\bibfnamefont {A.~E.}\ \bibnamefont
  {Allahverdyan}}, \bibinfo {author} {\bibfnamefont {R.}~\bibnamefont
  {Balian}}, \ and\ \bibinfo {author} {\bibfnamefont {T.~M.}\ \bibnamefont
  {Nieuwenhuizen}},\ }\href {\doibase 10.1209/epl/i2004-10101-2} {\bibfield
  {journal} {\bibinfo  {journal} {EPL (Europhys. Lett.)}\ }\textbf {\bibinfo
  {volume} {67}},\ \bibinfo {pages} {565} (\bibinfo {year} {2004})}\BibitemShut
  {NoStop}%
\bibitem [{\citenamefont {Leonhardt}(1997)}]{Leonhardt_BOOK}%
  \BibitemOpen
  \bibfield  {author} {\bibinfo {author} {\bibfnamefont {U.}~\bibnamefont
  {Leonhardt}},\ }\href {https://books.google.ca/books?id=wmsJy1A\_cyIC} {\emph
  {\bibinfo {title} {Measuring the Quantum State of Light}}},\ Cambridge
  Studies in Modern Optics\ (\bibinfo  {publisher} {Cambridge University
  Press},\ \bibinfo {year} {1997})\BibitemShut {NoStop}%
\bibitem [{\citenamefont {Skrzypczyk}, \citenamefont {Silva},\ and\
  \citenamefont {Brunner}(2015)}]{Skrzypczyk_2015}%
  \BibitemOpen
  \bibfield  {author} {\bibinfo {author} {\bibfnamefont {P.}~\bibnamefont
  {Skrzypczyk}}, \bibinfo {author} {\bibfnamefont {R.}~\bibnamefont {Silva}}, \
  and\ \bibinfo {author} {\bibfnamefont {N.}~\bibnamefont {Brunner}},\ }\href
  {\doibase 10.1103/PhysRevE.91.052133} {\bibfield  {journal} {\bibinfo
  {journal} {Phys. Rev. E}\ }\textbf {\bibinfo {volume} {91}},\ \bibinfo
  {pages} {052133} (\bibinfo {year} {2015})}\BibitemShut {NoStop}%
\bibitem [{\citenamefont {Paris}(1996)}]{Paris_1996}%
  \BibitemOpen
  \bibfield  {author} {\bibinfo {author} {\bibfnamefont {M.~G.}\ \bibnamefont
  {Paris}},\ }\href {\doibase 10.1016/0375-9601(96)00339-8} {\bibfield
  {journal} {\bibinfo  {journal} {Phys. Lett. A}\ }\textbf {\bibinfo {volume}
  {217}},\ \bibinfo {pages} {78 } (\bibinfo {year} {1996})}\BibitemShut
  {NoStop}%
\bibitem [{\citenamefont {Lvovsky}\ and\ \citenamefont
  {Babichev}(2002)}]{Lvovsky_2002}%
  \BibitemOpen
  \bibfield  {author} {\bibinfo {author} {\bibfnamefont {A.~I.}\ \bibnamefont
  {Lvovsky}}\ and\ \bibinfo {author} {\bibfnamefont {S.~A.}\ \bibnamefont
  {Babichev}},\ }\href {\doibase 10.1103/PhysRevA.66.011801} {\bibfield
  {journal} {\bibinfo  {journal} {Phys. Rev. A}\ }\textbf {\bibinfo {volume}
  {66}},\ \bibinfo {pages} {011801} (\bibinfo {year} {2002})}\BibitemShut
  {NoStop}%
\bibitem [{\citenamefont {Oxborrow}, \citenamefont {Breeze},\ and\
  \citenamefont {Alford}(2012)}]{Oxborrow_2012}%
  \BibitemOpen
  \bibfield  {author} {\bibinfo {author} {\bibfnamefont {M.}~\bibnamefont
  {Oxborrow}}, \bibinfo {author} {\bibfnamefont {J.~D.}\ \bibnamefont
  {Breeze}}, \ and\ \bibinfo {author} {\bibfnamefont {N.~M.}\ \bibnamefont
  {Alford}},\ }\href {\doibase 10.1038/nature11339} {\bibfield  {journal}
  {\bibinfo  {journal} {Nature}\ }\textbf {\bibinfo {volume} {488}},\ \bibinfo
  {pages} {353} (\bibinfo {year} {2012})}\BibitemShut {NoStop}%
\bibitem [{\citenamefont {Singer}(2013)}]{Singer_BOOK}%
  \BibitemOpen
  \bibfield  {author} {\bibinfo {author} {\bibfnamefont {J.}~\bibnamefont
  {Singer}},\ }\href {https://books.google.co.il/books?id=ghpyngEACAAJ} {\emph
  {\bibinfo {title} {Masers}}}\ (\bibinfo  {publisher} {Literary Licensing,
  LLC},\ \bibinfo {year} {2013})\BibitemShut {NoStop}%
\bibitem [{\citenamefont {Gardiner}\ and\ \citenamefont
  {Zoller}(2000)}]{Gardiner_BOOK}%
  \BibitemOpen
  \bibfield  {author} {\bibinfo {author} {\bibfnamefont {C.~W.}\ \bibnamefont
  {Gardiner}}\ and\ \bibinfo {author} {\bibfnamefont {P.}~\bibnamefont
  {Zoller}},\ }\href@noop {} {\emph {\bibinfo {title} {Quantum Noise}}}\
  (\bibinfo  {publisher} {Springer},\ \bibinfo {address} {Berlin},\ \bibinfo
  {year} {2000})\BibitemShut {NoStop}%
\bibitem [{\citenamefont {Walls}\ and\ \citenamefont
  {Milburn}(1994)}]{Walls_BOOK}%
  \BibitemOpen
  \bibfield  {author} {\bibinfo {author} {\bibfnamefont {D.~F.}\ \bibnamefont
  {Walls}}\ and\ \bibinfo {author} {\bibfnamefont {G.~J.}\ \bibnamefont
  {Milburn}},\ }\href@noop {} {\emph {\bibinfo {title} {Quantum Optics}}},\
  \bibinfo {edition} {1st}\ ed.\ (\bibinfo  {publisher} {Springer-Verlag},\
  \bibinfo {address} {Berlin},\ \bibinfo {year} {1994})\BibitemShut {NoStop}%
\bibitem [{\citenamefont {Ghosh}\ \emph {et~al.}(2017)\citenamefont {Ghosh},
  \citenamefont {Latune}, \citenamefont {Davidovich},\ and\ \citenamefont
  {Kurizki}}]{Ghosh_2017}%
  \BibitemOpen
  \bibfield  {author} {\bibinfo {author} {\bibfnamefont {A.}~\bibnamefont
  {Ghosh}}, \bibinfo {author} {\bibfnamefont {C.~L.}\ \bibnamefont {Latune}},
  \bibinfo {author} {\bibfnamefont {L.}~\bibnamefont {Davidovich}}, \ and\
  \bibinfo {author} {\bibfnamefont {G.}~\bibnamefont {Kurizki}},\ }\href
  {\doibase 10.1073/pnas.1711381114} {\bibfield  {journal} {\bibinfo  {journal}
  {Proceedings of the National Academy of Sciences}\ }\textbf {\bibinfo
  {volume} {114}},\ \bibinfo {pages} {12156} (\bibinfo {year}
  {2017})}\BibitemShut {NoStop}%
\end{thebibliography}

%

\appendix

\section{Derivation of the Raman Hamiltonian}\label{Suppl-A}

Let us denote the hot-bath states with frequencies $\omega_{k} \simeq \omega_h$ and wave vectors $k$ by their occupation numbers $n_{k}$ and the signal-mode with occupation $n_s$ by $|n_s\rangle$. If we have $|g,n_{k},n_s\rangle=|\Psi_i\rangle$ as the initial and $|e,n_{k}-1,n_s+1\rangle=|\Psi_f\rangle$ as the final states, we will have experienced a Raman process (Fig.~\ref{RamanFig}). According to time-dependent perturbation theory, the probability amplitude for a transition $|\Psi_i\rangle$ to $|\Psi_f\rangle$ after a time $t$ is given by
	\begin{widetext}
\begin{gather}
\mathcal{A}_{i\rightarrow f}=\langle \Psi_f|\left(1+\frac{1}{i\hbar}\int^{t}_{0}dt_1 V(t_1)+\frac{1}{(i\hbar)^{2}}\int^{t}_{0}dt_1\int^{t_1}_{0}dt_2 V(t_1)V(t_2)+...\right)|\Psi_i\rangle,\nonumber\\
\simeq -\frac{1}{\hbar^{2}}\int^{t}_{0}dt_1\int^{t'}_{0}dt_2 \langle \Psi_f|V(t_1)V(t_2)|\Psi_i\rangle,
\label{Perturbation-theory}
\end{gather}
since $\langle \Psi_f|\Psi_i\rangle=\langle \Psi_f|V|\Psi_i\rangle=0$. We keep only the rotating-wave term and denote a virtual level that enables the Raman transition by $|u\rangle$ and its energy by  $\epsilon_u$, and the corresponding dipolar couplings by $g^{(\omega_{k})}_{ug}$ and $g^{(\nu)}_{ue}$ respectively. The second-order probability amplitude of the two-photon transition $|\Psi_i\rangle=|g,n_{k},n_s\rangle \Rightarrow |u,n_{k}-1,n_s\rangle  \Rightarrow |\Psi_f\rangle=|e,n_{k}-1,n_s+1\rangle$ is then
\begin{gather}
-\mathcal{A}_{i\rightarrow f}
\simeq \sum_{k}\int^{t}_{0}dt_1\int^{t_1}_{0}dt_2 g^{(\nu)}_{eu}g^{(\omega_{k})}_{ug}\langle n_{k}-1,n_s+1|e^{it_1(\epsilon_e-\epsilon_u)}{\bop}^{\dag}e^{i\nu t_1}e^{it_2(\epsilon_u-\epsilon_g)}a_{k}e^{-i\omega_{k} t_2} |n_{k},n_s\rangle.
\label{prob-amplitude}
\end{gather}

\par

\begin{figure}
	\centering
		\includegraphics[width=\columnwidth]{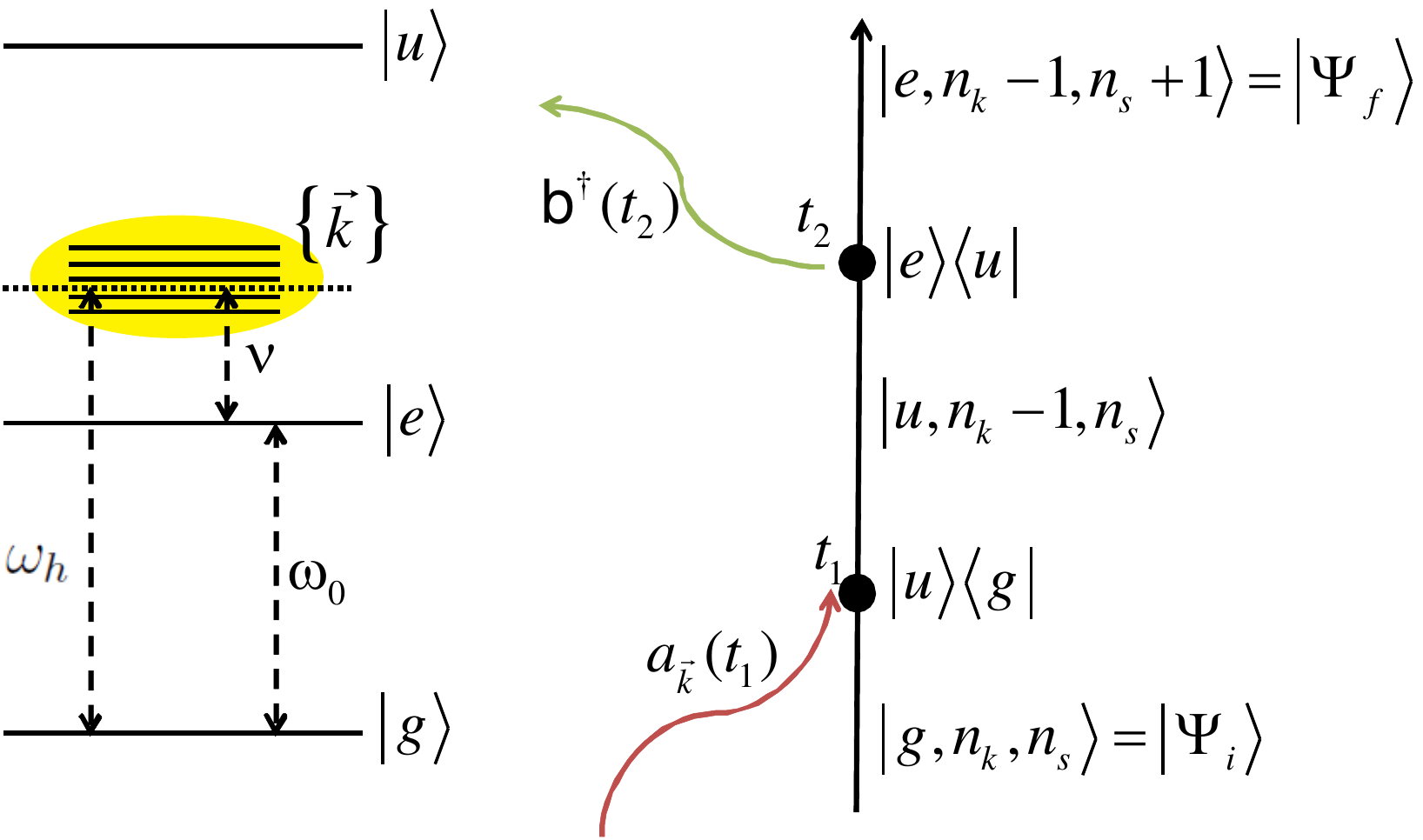}
\caption{Illustration to Appendix \ref{Suppl-A}}  \label{RamanFig}
\end{figure}

\par

Assuming that the level energies satisfy $\epsilon_u>\epsilon_e> \epsilon_g$ and $0< \nu < \omega_{k}$, we take the frequencies to be non-resonant (in the one-photon sense). Thus we evaluate the amplitude of the rotating term to be
\begin{gather}
-\mathcal{A}_{i\rightarrow f} 
=\sum_{k} g^{(\nu)}_{eu}g^{(\omega_{k})}_{ug}\frac{-i\sqrt{n_{k}}\sqrt{n_s+1}}{i(\epsilon_u-\epsilon_g-\omega_{k})}
\left\{\frac{e^{it[(\epsilon_e-\epsilon_g)-(\omega_{k}-\nu)]}-1}{(\epsilon_e-\epsilon_g)-(\omega_{k}-\nu)}
-\frac{e^{it(\epsilon_e-\epsilon_u+\nu)}-1}{\epsilon_e-\epsilon_u+\nu}\right\},
\label{term-R}
\end{gather}
which, in the limit $t\rightarrow \infty$ and upon neglecting the principle-value terms, reduces to
\begin{gather}
-\mathcal{A}_{i\rightarrow f}=
\sum_{k} g^{(\nu)}_{eu}g^{(\omega_{k})}_{ug}(-2\pi i)\frac{\delta[(\epsilon_e-\epsilon_g)-(\omega_{k}-\nu)]}{(\epsilon_u-\epsilon_g-\omega_{k})}\sqrt{n_{k}}\sqrt{n_s+1}.
\label{term-R-reduce}
\end{gather}
The Raman Hamiltonian should yield the same transition amplitude from first-order perturbation theory 
\begin{gather}
\mathcal{A}_{i\rightarrow f}=\langle \Psi_f|\left(1+\frac{1}{i\hbar}\int^{t}_{0}dt_1 V_h(t_1)\right)|\Psi_i\rangle.
\label{1st-order-perturbation-theory}
\end{gather}
Therefore, the Raman Hamiltonian is
\begin{gather}
V_h(t)={2\pi}{\hbar}\sum_{k}\frac{g^{(\nu)}_{eu}g^{(\omega_{k})}_{ug}}{\epsilon_u-\epsilon_g-\omega_{k}}{\bop}^{\dag} a_k e^{-i(\omega_{k}-\nu)t}|e\rangle\langle g| e^{i(\epsilon_e-\epsilon_g)t}+{\rm H.c.},
\label{1st-order-perturbation-theory}
\end{gather}
which is equivalent to 
\begin{eqnarray}
V_h(t)&=&\hbar\sum_{k}g^{(h)}_{k} \left(|g\rangle\langle e|{\bop}^\dag a^{(h)}_{k} e^{-i[\omega_{k}-(\nu+\omega_0)]t}+ {\rm H. c.}\right),
\label{V(t)-reduce}
\end{eqnarray}
with the coupling constant
\begin{equation}\label{gkh}	g_{ k}^{(h)}=2\pi\frac{g^{(\nu)}_{eu}g^{(\omega_{k})}_{ug}}{\epsilon_u-\epsilon_g-\omega_{h}}, 
\end{equation}
taken to be real and $\omega_k\approx\omega_h$, approximated near the two-photon Raman resonance. 

\par

Our next goal is to obtain the master equation of motion for the density operator $\rho$ of the joint (WM+signal) system coupled to the hot bath:
\begin{gather}
\dot{\rho}^{(h)}=-\frac{i}{\hbar}{\rm Tr}_{h}[V(t),\rho(t_0)\otimes \rho_{h}(t_0)]
-\frac{1}{\hbar^{2}}{\rm Tr}_{h}\int^{t}_{t_0}[V(t),[V(t'),\rho(t')\otimes \rho_{h}(t_0)]dt'.
\label{rho-dot}
\end{gather}
\end{widetext}
Inserting $V(t)$ into Eq.~\eqref{rho-dot}, we recall that $\langle a_{k} \rangle= \langle a^{\dag}_{k} \rangle=0$ and $\langle a_{k} a_{{k}'} \rangle= \langle a^{\dag}_{k} a^{\dag}_{{k}'} \rangle=0$, where $\langle A \rangle={\rm Tr}_h[\rho_h(t_0)A]$ stands for tracing over the hot bath and we omit the superscript $(h)$. Next we note that
\begin{gather}
\langle a^{\dag}_{k} a_{{k}'} \rangle= \bar{n}_{k} \delta_{kk'}; \quad \langle  a_{k} a^{\dag}_{{k}'}\rangle= (\bar{n}_{k}+1) \delta_{kk'}.
\label{hot-bath-commutation-rel}
\end{gather}
Then sum over $k$ can be replaced by an integral as follows:
\begin{gather}
\sum_{k} \rightarrow \frac{\mathcal{V}}{\pi^{2}}\int^{\infty}_{0} d^{3}k,
\label{sum-to-integral}
\end{gather}
where $\mathcal{V}$ is the hot-bath mode volume. Hereafter we neglect all memory effects (i.e.,  we adopt the Markov approximation) and assume that $\rho(t')$ is slowly varying. We can thus extend the integration over $t'$ to $\infty$ and use
\begin{gather}
\int d^{3}k\int^{\infty}_{t_0} dt' e^{i(\omega-c k)(t-t')} \simeq L(\omega-\omega_h),
\label{sum-to-integral}
\end{gather}
where $L(\omega-\omega_h)$ is a Lorentzian shape whose (narrow) width is determined by the cavity-mirror finesse, and
\begin{gather}
\frac{\omega_0+\nu}{c}=\frac{\omega_h}{c}.
\label{k_0}
\end{gather}
We then obtain from \eqref{rho-dot}
\begin{gather}
\dot{\rho}^{(h)}=\gamma_h(\bar{n}_h+1)([S\rho,S^{\dag}]+[S,\rho S^{\dag}])\nonumber\\
+\gamma_h\bar{n}_h([S^{\dag}\rho,S]+[S^{\dag},\rho S]),
\label{final-QME-rho-dot-reduce}
\end{gather}
where 
\begin{equation}\label{Sop}
S=  {\bop}|g\rangle\langle e|\equiv {\bop}\sigma_- ,
\end{equation}
and
\begin{gather}
\gamma_h=\frac{\mathcal{V}\omega^{2}_{h}g^{2}_{k}}{\pi c^{3}},
\label{gamma_h}
\end{gather}
is the decay rate into the hot bath calculated for the Raman coupling $g^{2}_{k}$.
The equation of motion (\ref{final-QME-rho-dot-reduce}) is identical to the standard master equation for a two-level system coupled to a thermal reservoir \cite{Scully_BOOK}, which in our case governs the interaction of the WM with the cold bath:
\begin{gather}
\dot{\rho}^{(c)}=\gamma_c(\bar{n}_c+1)([\sigma_-\rho,\sigma_+]+[\sigma_+,\rho \sigma_+])\nonumber\\
+\gamma_c\bar{n}_c([\sigma_+\rho,\sigma_-]+[\sigma_+,\rho \sigma_-]).
\label{final-QME-rho-dot-reduce-cold}
\end{gather}
via the Hamiltonian 
\begin{eqnarray}
V_c(t)&=&\hbar\sum_{k}g^{(c)}_{k} \left(|g\rangle\langle e|a^{(c)}_{k} e^{-i[\omega_{k}-\omega_0]t}+ {\rm H. c.}\right),\nonumber\\
\label{V(t)-reduce0}
\end{eqnarray}
where the summation is over bath wavevectors $k$, and $g^{(c)}_{k}$ is the coupling constants, taken to be real. However, in the case of the hot bath, the transition operator 
$ S=  {\bop}|g\rangle\langle e|$ involves both the WM and the signal. Futhermore, the hot bath coupling constant, arising from Eqs.~(\ref{gkh}) and (\ref{gamma_h}), is typically much weaker than its cold bath counterpart. Taken together, master equations (\ref{final-QME-rho-dot-reduce}) and (\ref{final-QME-rho-dot-reduce-cold}) govern the dynamics of the WM+signal system.

\section{Evolution of the signal state}\label{Suppl-B}

Since the $|g\rangle \leftrightarrow |e\rangle$ transition is primarily caused by the cold-bath transition at a rate $\gamma_c\gg\gamma_h$, we can assume that the WM attains its steady state in thermal resonance with the cold bath, so Eq.~(\ref{population-ratio}) holds. Upon tracing out the  WM, we obtain the reduced (Lindblad) master equation for the signal-mode density matrix  $\rho_s=\operatorname{Tr}_W(\rho)$
\begin{gather}
\dot{\rho}_s=\gamma_h(\bar{n}_{h}+1)\rho_{ee}([{\bop}\rho_s,{\bop}^{\dag}]+[{\bop},\rho_s {\bop}^{\dag}])\nonumber\\
+\gamma_h\bar{n}_{h}\rho_{gg}([{\bop}^{\dag}\rho_s,{\bop}]+[{\bop}^{\dag},\rho_s {\bop}]),\label{rho-s-dot}
\end{gather}
and therefore
\begin{gather}
\dot{\bar{n}}_s=-\gamma_h\left[\rho_{ee}\bar{n}_s(\bar{n}_{h}+1)-\rho_{gg}\bar{n}_h(\bar{n}_s+1)\right]. 
\label{n-s-dot}
\end{gather}
In the semiclassical limit, setting $\bar{n}_s=I_s$ and ignoring spontaneous emission into the hot bath, we get Eqs.~(\ref{rateeq-scully}) and (\ref{rateeq-scully-mod}).

\par

To treat the signal quantum-mechanically, we use the Glauber-Sudarshan decomposition of the signal state:
\begin{gather}
\rho_s=\int{d^2 \alpha P(\alpha) |\alpha \rangle \langle \alpha|},\label{coherent-state-expansion}
\end{gather} 
where $P(\alpha)$ is the Glauber-Sudarshan $P$ quasiprobability distribution. We then obtain the Fokker-Planck (FP) equation for this distribution \cite{Scully_BOOK,Gardiner_BOOK,Walls_BOOK}
\begin{gather}
\frac{\partial P}{\partial t}=-\frac{G}{2}\left(\frac{\partial}{\partial\alpha}
+\frac{\partial}{\partial\alpha*}\right)P+D\frac{\partial^{2}P}{\partial\alpha\partial\alpha*},
\end{gather}
with
\begin{gather}
G =\frac{\gamma_h\bar{n}_h(\bar{n}_{c}+1)-\gamma_h(\bar{n}_{h}+1)\bar{n}_{c}}{2\bar{n}_{c}+1}
= \frac{\gamma_h(\bar{n}_h-\bar{n}_{c})}{2\bar{n}_{c}+1};\\
D=\frac{\gamma_h\bar{n}_h(\bar{n}_{c}+1)}{2\bar{n}_{c}+1}.\label{eq:FP-sqz}
\end{gather}
Here $G$ is the effective gain rate in the amplification regime [which is identical to the semiclassical one \eqref{rateeq-scully}] and $D$ is the diffusion rate. For a coherent state input, this corresponds \cite{Scully_BOOK} to the $P$ function behaving in accordance with Eq.~\eqref{Pevolve} with $\sigma^2(t)=\frac{D}{G}(e^{G t}-1).$

\par

\section{Implementation of the displacement operator}\label{Suppl-C}

The beam splitter will enact the following transformation:
\begin{align}
a_\mathrm{LO}&\to r a_\mathrm{LO}+\tau a;\nonumber\\
a&\to r a - \tau a_\mathrm{LO}.
\end{align}
Here $a_\mathrm{LO}$ is the amplitude operator of the local oscillator field (which can be treated as a c-number because of its high magnitude compared to $\alpha_0 e^{G t/2}$), and $\tau \ll 1$ and $r$ are the beam splitter transmissivity and reflectivity, respectively. By setting $\tau a_\mathrm{LO}=r\alpha_0e^{G t/2}$, we obtain a state of zero amplitude in the reflected channel of the signal. The energy of the local oscillator, on the other hand, will increase by the coherent-component energy according to 
\begin{align}
\hbar\nu a_\mathrm{LO}^2&\to \hbar\nu[r a_\mathrm{LO}+\tau \alpha_0e^{G t/2}]^2\\
&\approx
\hbar\nu[(1-\tau^2/2) a_\mathrm{LO}+\tau^2 a_\mathrm{LO}]^2\\
&\approx\hbar\nu a_\mathrm{LO}^2(1+\tau^2)\\
&\approx\hbar\nu a_\mathrm{LO}^2+[\alpha_0e^{G t/2}]^2,
\end{align}
where we have approximated 
\begin{align}
r=\sqrt{1-\tau^2}\approx 1-\tau^2/2.
\end{align}

\section{Derivation of the efficiency Eq. \eqref{eff}}\label{Suppl-D}

The efficiency can be computed as the ratio
\begin{gather}
\eta=\frac{\dot{\mathcal{W}}_s}{\dot{Q}_h},
\label{angular-component}
\end{gather}
where $\dot{\mathcal{W}}_s=\hbar\nu G|\alpha_{0}|^{2}e^{G t}$ and the heat flux from the hot bath is given by
$\dot{Q}_h=\omega_h[G\langle \bop^{\dag}\bop \rangle + D]$ \cite{Ghosh_2017}.

\par

Using Eq. \eqref{eq:Hstavg}, $\langle H_s(t)\rangle=\hbar\nu \langle \bop^{\dag}\bop \rangle$, the expression for the efficiency can be simplified as
\begin{equation}\label{eff-expression}
\eta=\frac{\nu}{\omega_{h}}\frac{|\alpha_{0}|^{2}}{|\alpha_{0}|^{2}+\frac{D}{G}}.
\end{equation}
Equation \eqref{eff-expression} can be further rewritten as [Cf. \eqref{eff}]: 
\begin{equation}\label{eff-alternative}
\eta=\eta_{SSD}\frac{\hbar\nu|\alpha_{0}|^{2}e^{G t}-\langle H_s(0)\rangle}{\langle H_s(t)\rangle-\langle H_s(0)\rangle}=\frac{\nu}{\omega_{h}}\frac{|\alpha_{0}|^{2}}{|\alpha_{0}|^{2}+\frac{\bar{n}_h(\bar{n}_{c}+1)}{\bar{n}_{h}-\bar{n}_{c}}},
\end{equation}
where we take into account that the first term of the numerator \eqref{eff-alternative} corresponds to extractable work and the ergotropy of the initial coherent state is $\langle H_s(0)\rangle=\hbar\nu|\alpha_{0}|^{2}$.

\end{document}